\documentstyle[ltwol]{article}

\def\Journal#1#2#3#4{{#1} {\bf #2}, #3 (#4)}

\def\NPB{{\em Nucl. Phys.} B}
\def\PLB{{\em Phys. Lett.}  B}
\def\PRL{\em Phys. Rev. Lett.}
\def\PRD{{\em Phys. Rev.} D}

\def\CMP{{\em Commun. Math. Phys.}}

\def\f#1#2{\textstyle{#1\over #2}}

\def\ev#1{\langle#1\rangle}

\def\be{\begin{equation}}
\def\ee{\end{equation}}
\def\bea{\begin{eqnarray}}
\def\eea{\end{eqnarray}}

\begin{document}

\title{``NON-PERTURBATIVE METHODS'' IN FIELD THEORY}

\author{KENNETH INTRILIGATOR}

\address{UCSD Physics Department, 9500 Gilman Drive, La Jolla CA
92093, USA}   


\twocolumn[\maketitle\abstracts{This talk is an overview of selected
topics related to renormalization group flows and the phases of gauge 
theories.}]

\section{Introduction}

In an asymptotically free gauge theory, starting at small coupling
$g$, the coupling $g(\mu /\Lambda)\rightarrow 0$ in the ultra-violet
and perturbation theory is valid.  In the infra-red, on the other
hand, $g(\mu /\Lambda)$ becomes large and we might wonder if there are
some ``non-perturbative methods'' (which was the assigned title of my
talk) which can be applied to the problem.
The analysis in the IR looks hard, but often that's only because we're
making a mistake in trying to describe IR physics in terms of the UV
variables.  The physics of the IR is often better described using a
weakly coupled, effective field theory for the light degrees of
freedom, for example the chiral lagrangian for pions.  

The question to ask, then, is {\it ``what are the correct variables
and interactions in the IR?''}  There are two paths for answering
this question.  The first is to {\it derive} the answer directly from
the ultraviolet lagrangian using some sort of non-perturbative
methods.  This path is extremely hard.  While there have been
important developments in such non-perturbative methods, there does
not seem to be a systematic way to determine even when such methods
are or are not applicable, even for answering only qualitative
questions about the physics.  (This possibly partly reflects
my own ignorance concerning this path.)  

The second path to answering the above question is to use symmetries,
match to known results, use guess-work if needed, do some non-trivial
cross-checks, then conjecture you've ``solved'' the theory!  In other
words, in this path you cheat, getting to the answer without doing all
the hard work.  This path has been very fruitful over the past few
years (and longer, e.g. it's the path from QCD to pions and the chiral
lagrangian).  In this talk, I will mostly discuss results obtained by
this second path, especially using supersymmetry.  Of course, in the
long-term, we would also like to have non-perturbative methods which
are powerful enough to be able to {\it derive} these results directly
from the UV lagrangian.  In any case, knowing the answers should prove
useful in developing more direct non-perturbative methods.

\section{Renormalization group flows to the IR}

Theories can flow under the renormalization group to either a free or
interacting, scale invariant, RG fixed point in the extreme IR.  We
can schematically picture the flows in theory space, with coordinates
given by the various coupling constants.  Intuitively, we can picture
the RG flows as streams of water, flowing over mountains and through
valleys, into lakes.  The ``lakes,'' which are the fixed points, can
be either free or interacting.  Each fixed point is the end point of
flows for all theories with coupling constants $g_i$ in some
particular basin of attraction.  The basins of attraction do not
overlap: a given theory will flow to a unique endpoint.  

There are some flows which, when plotted in theory space, with the
coupling constant coordinates, look funny. For example, while one flow
ends up in one fixed point, another flow, which starts off parallel
and near to the first flow, can end up veering away from the first
flow and eventually flow into a different fixed point, possibly quite
distant from that of the first flow.  In other words, seemingly nearby
points can be in very different basins of attraction.  In the picture
of streams of water, these funny looking flows are due to various
mountain ridges not shown on the map.  While the two flows described
above initially seemed nearby, they were actually separated by a large
mountain ridge and thus wound up flowing through different valleys and
into different lakes.

The intuition behind the above picture of the renormalization group
flow is that massive degrees of freedom decouple in RG flows to the
infrared -- that there is a thinning of degrees of freedom.  Because of
this, the RG flow is {\it irreversible}.  The flow can not circle around
back to where it was before, and thus there are no limit cycles where
the flow forever circles around in a closed loop.  As with the steams
of water, the intuition is that the RG always flows ``downward.'' 

This was proven in 2d by Zamolodchikov~\cite{Zamo} for any unitary
theory. He showed that there is a ``c-function,'' $c(g_i)$, which
monotonically decreases along all RG flows and is stationary at the RG
fixed points.  The beta functions $\beta _i=dg_i/d(log \mu)$, which
give the ``velocity vector,'' $v_i=-\beta _i$, of the flow to the IR,
are the gradients of the c-function: $\beta _i=\partial c/\partial
g_i$.  The c-function $c(g_i)$, which counts the number of degrees of
freedom of the theory, thus corresponds to the notion of height in the
picture of flowing water.  It was also shown~\cite{Zamo} that there is
a positive definite metric $G^{ij}(g)$ on theory space, which can be
used to measure distances between theories.  The metric allows the
funny flows to be understood, as it gives the information about if
there are mountains.  Two points $g_i^{(1)}$ and $g_i^{(2)}$ which
seem nearby on the map are actually separated by a mountain ridge if
$\int _{g_i^{(1)}}^{g_i^{(2)}}\sqrt{G^{kl}dg_kdg_l}$ is always large.
The RG flows, as with water flows, are minimum distance, geodesics
with respect to the metric $G^{ij}$.

We expect a similar situation in 4d, though there the {\it proof} has
been elusive.  A candidate c function, analogous to the 2d case, has
been conjectured by Cardy~\cite{Cardy}
$$c(g)={120\over \pi ^2}\int _{S^4}\ev{T^\mu _\mu}\sqrt{g}d^4x.$$ For
free fields, this $c=124N_1+11N_{1/2}+2N_0$, where $N_j$ is the number
of fields of spin $j$.  The question, then, is if this function indeed
monotonically decreases along all RG flows.  A weaker statement to
check is if the UV and IR endpoints of flows satisfy $c_{UV}>c_{IR}$.
The fact that $c_{UV}>c_{IR}$ for Cardy's c-function was been checked
in a number of ``solved'' supersymmetric gauge theory
examples~\cite{Anselm}, with various other candidate c-functions ruled
out.  A recent claim is that this longstanding problem has finally
been solved and that Cardy's $c(g_i)$ can indeed be proven to
monotonically decrease along all RG flows for any unitary
theory~\cite{Forte}.  It remains to be seen if the arguments
of~\cite{Forte}, which has been regarded with some scepticism by
some~\cite{anom}, are really an airtight proof of the c-theorem.  In
any case, it seems likely that there is a c-theorem in 4d and that
this is the correct c-function.

\section{Anomaly matching}

A useful constraint on where RG flows possibly end up are the 't Hooft
anomaly matching conditions~\cite{Hooft}.  The 't Hooft anomalies are
the obstructions, $Tr H$ and $Tr H^3$, to gauging a global symmetry
$H$.  For unbroken $H$, these anomalies can be evaluated knowing only
the massless fermion spectrum.  't Hooft argued that these quantities
are {\it constant along RG flows}.  Thus the original UV theory and
the IR fixed point must have the same 't Hooft anomalies.  Also, all
theories which flow to the same RG fixed point must have the same 't
Hooft anomalies.  This is a useful constraint for {\it ruling out}
scenarios about where theories flow: if the 't Hooft anomalies don't
match, it's wrong!

It was recently argued~\cite{CsaMur} that anomaly matching for
discrete symmetries is also a useful constraint for ruling out various
scenarios.  In particular, it was argued there that the recently
conjectured~\cite{KovShif} ``chirally symmetric phase'' of
${\cal N}=1$ super Yang-Mills, which will be discussed further in
sect. 6, can be ruled out on the basis of matching anomalies for the
discrete $Z_{2h}\subset U(1)_R$ left unbroken by instantons.  This was
criticized in~\cite{KKS} on the basis that, if the discrete
symmetry is the remnant of a spontaneously broken continuous symmetry,
there will be Goldstone bosons present which ensure that the discrete
anomalies are always matched.  However, it is not clear why this
criticism should be applicable for theories, such as the one under
consideration, where the discrete symmetry does not come from
spontaneously breaking a continuous symmetry and there is is no
Goldstone boson present.

Anomaly matching which {\it does} work for a given scenario, in a way
which is non-trivial, can be regarded as non-trivial evidence that the
scenario is correct.  For example, confinement was thus
argued~\cite{ISS} to occur in ${\cal N}=1$ supersymmetric $SU(2)$
gauge theory with a single matter field $Q$ in the ${\bf 4}$
representation of $SU(2)$.  

There are, however, some cautions to point out regarding anomaly
matching.  One is that global symmetries of one theory might not be
manifest in another which flows to the same fixed point.  This
phenomenon, where a theory has a larger global symmetry in the
infrared, is that of ``accidental global symmetries.'' This point was
emphasized in~\cite{AccidLS} and illustrated in a variety of
supersymmetric examples.  Another caution is that there are known
examples of numerically miraculous, but physically misleading
matching, suggesting confinement, in a class of models which are
argued to definitely not confine but, rather, have interacting RG
fixed points~\cite{BCI}.  For example, this is the case for ${\cal
N}=1$ supersymmetric $SO(N)$ with a single matter chiral superfield in
the two-index symmetric tensor representation of $SO(N)$.  There is a
highly non-trivial anomaly matching, which holds for all $N$,
suggesting confinement.  Nevertheless, the theories actually do not
confine.

\section{Interacting RG fixed points}

An interacting RG fixed point is a non-trivial conformal field theory.
While there are many known conformal field theories in 2d, they were
previously considered to be quite rare and exotic in 4d.  A surprise
which has been learned from studying supersymmetric theories is that
they are actually very common!  Indeed, they {\it generically occur if
there is enough matter}.  This is not special to supersymmetry. A
basic scenario for having a RG fixed point dates back to~\cite{DGFW}.
Suppose the matter content is such that the one-loop beta function is
negative (asymptotically free in UV) while the two-loop beta function
is positive: $\beta (g)=-b_1g^3+b_2g^5+\dots$.  Then $\beta
(g^*\approx\sqrt{b_1/b_2})=0$ and perturbation theory suggests that
the theory has a fixed point. As long as $g^*=\sqrt{b_1/b_2}$ is
small, one is inclined to trust perturbation theory and believe that
the fixed point actually exists.

For ordinary (non-supersymmetric) $SU(N_c)$ QCD with $N_f$ flavors of
quarks in the ${\bf N_c+\overline N_c}$, the above requirement on the
signs of the one and two-loop contributions to the beta function are
satisfied for $N_f$ just below $11N_c/2$.  For example, for
$N_f={11\over 2}N_c-1$ we have $b_1\sim 1$ and $b_2\sim N_c^2$ and thus
expect a RG fixed point with $g^*\sim 1/N_c$.  For large $N_c$ we can
trust perturbation theory (the scaled coupling $(g^*)^2N_c\sim 1/N_c$
is also small) and are thus inclined to believe that the RG fixed
point really exists~\cite{BZ}. For $N_f$ any larger, $N_f\geq
11N_c/2$, the theory ceases to be asymptotically free and flows to a
free theory in the infrared.

Decreasing $N_f$ below $N_f={11\over 2}N_c-1$, the value of $g^*$,
where the perturbative beta function suggests a fixed point, tends to
increase.  For low enough $N_f$, we are then less inclined to believe
perturbation theory and need non-perturbative methods to determine
whether or not the theory continues to have a RG fixed point.  The
expectation is that for some entire range of flavors, ${11\over
2}N_c>N_f>N_f^*$, the theory has RG fixed points. Then, at some
critical number of flavors, $N_f^*$, it goes over to a new phase.  
Eventually, for low enough $N_f$, the theory is expected to be in a
confining phase.  

As will be reviewed later, this flavor dependent phase structure has
been well studied in the supersymmetric context using method two:
symmetries and non-trivial cross checks.  

For the present case of non-supersymmetric QCD, the phase structure
has been analyzed using method one, with a variety of more direct
approximations for the non-perturbative regime, both on the lattice
and in the continuum.  I will only mention here a result obtained in
the continuum.  In~\cite{ATW} the situation was considered where one
starts at a non-trivial RG fixed point with some number $N_f$ massless
flavors, and gives a mass to one flavor.  In the infrared, this theory
flows to QCD with $N_f-1$ massless flavors.  Analyzing such flows, it
was argued that there is a direct transition from a RG fixed point,
with no confinement or chiral symmetry breaking, to a phase with
confinement with chiral symmetry breaking at $N_f^*\approx 4N_c$.
This pattern differs from that seen in supersymmetric QCD
where~\cite{Nati}, decreasing $N_f$, the phases and transitions are
(RG fixed point) $\rightarrow$ (non-Abelian free magnetic phase)
$\rightarrow$ (confinement without chiral symmetry breaking)
$\rightarrow$ (confinement with chiral symmetry breaking).  There do
exist, however, other supersymmetric examples which exhibit a direct
(RG fixed point) $\rightarrow$ (confinement) similar to that argued to
occur in the non-supersymmetric case; an example is in ${\cal N}=1$
supersymmetric $SO(8)$ with matter in the spinor and vector
representations~\cite{PS}.

What is the physics of interacting RG fixed points?  Consider two test
charge quarks $q$ and $\overline q$ separated by a distance $R$.
Because the theory is scale invariant, the potential $V(R)=f(g^*)/R$,
rather than the linear potential $V\sim \sigma R$ of the confining
phase.  Naively the function $f(g^*)=(g^*)^2$.  The physics is a
non-Abelian version of the familar Coulomb phase of electrodynamics.

Because all fields are massless, there is no useful notion of an
S-matrix for asymptotically separated states.  Instead, the correct
observables are correlation functions of operators $\ev{\prod _i{\cal
O}_i (x_i)}$.  Generally, scale invariance implies conformal
invariance, and thus these correlation functions are restricted by the
conformal symmetry group.  In $d$-spacetime dimensions, the additional
generators of conformal symmetries extend the Lorentz group
$SO(d-1,1)$ to $SO(d,2)$.  In particular, in 4d the conformal group is
$SO(4,2)\cong SU(2,2)$.  

The operators must form representations of the conformal group.  Each
representation is generally infinite dimensional and is given by a
primary operators and its descendents, which are given by space-time
derivatives of the primary operator.  The conformal symmetry group
restricts the correlation functions, e.g. a general two-point function
of primary operators is
$$\ev{{\cal O}_i(x_i){\cal O}_j(x_j)}={c_{ij}\delta _{\Delta _i,
\Delta _j}\over |x-y|^{2\Delta _i}},$$ with $c_{ij}$ constants,
i.e. vanishing unless the dimensions $\Delta _i$ and $\Delta _j$ of
the two operators are equal.  The conformal algebra, along with
unitarity, places further constraints on the possible spectrum of
operator dimensions $\Delta _i$, e.g. for a spinless operator $\Delta
\geq 1$ with $\Delta =1$ if and only if the corresponding operator
satisfies free field equations of motion~\cite{Mack}.  

One could calculate the operator dimensions and operator product
expansions e.g. in Banks-Zaks type fixed points via perturbation
theory.  This has not been well studied in any 4d examples until
recently, for ${\cal N}=4$ supersymmetry and related theories.

\section{Supersymmetric Theories}

In supersymmetric theories, the restrictions of conformal invariance
become much more powerful.  The conformal group and supersymmetry must
combine into a single, super-conformal symmetry group.  The symmetry
group elements can be represented as a supermatrix  
$$\pmatrix{SO(d,2)&Q \cr Q^\dagger &J_R},$$ where $Q$ are the
fermionic supersymmetry generators, including additional ones
associated with the superconformal transformations, and $J_R$ are
bosonic $R$-symmetries which rotate the supercharges.  It is possible
to show that such a supergroup, with $Q$ in the spinor representation
of $SO(d,2)$, can exist only for $d\leq 6$ ~\cite{Nahm,Minwalla}, so
superconformal theories are impossible above $d=6$.  The spectrum of
operator dimensions are also constrained by the superconformal
symmetry, e.g. it is possible to show that all operators satisfy
$$\Delta \geq {d-1\over 2}|q_R|,$$
where $\Delta$ is the dimension of the operator and $q_R$ is its
charge under a $U(1)$ subgroup of the $R$-symmetry group.  

More generally, the reason why supersymmetric theories are often
easier to ``solve'' via the second method is that all light fields,
coupling constants, masses, even $\Lambda _{QCD}\sim (e^{-{8\pi
^2\over g^2}+i\theta})^{1/b_1}$ are {\it complex} (in $d=4$).  Various
quantities are {\it holomorphic} in the fields and coupling constants.
This is the ``power of holomorphy,'' found by N. Seiberg~\cite{holom}.
Since unbroken supersymmetry implies that $E_{vac}=0$ always, there
can be no first order phase transitions.  The only possible phase
transitions are second order, with some order parameter.  In addition,
supersymmetry implies that even the possible second order phase
transitions occur at isolated points in the complex plane, and can
thus always be avoided; there are no ``walls'' separating phases.
(The closest thing to a ``wall'' in a supersymmetric theory is the
curve of marginal stability~\cite{swi}, where otherwise stable (BPS)
states can decay.)

For this reason, it is possible to obtain some exact results by
matching to known results in various limits, e.g. weak coupling, and
then analytically continuing in the various fields and coupling
constants, to obtain the exact result everywhere.  Only certain
quantities can be obtained in this way -- not all aspects of the
theory are ``solved.''  But the solvable aspects concern the most
interesting questions: the infrared physics.  The exact results thus
give useful insight into the strongly coupled dynamics of
supersymmetric theories.

By using analytic continuation, with no phase transitions, in masses
$m$ or field expectation values $v$, along with decoupling arguments,
many results for different theories are interrelated.  There is thus a
growing web of interrelated results of different models, with many
cross-checks. See, for example~\cite{isrev} for a number of early
examples and references.  

The exact results for supersymmetric models have two types of
applications for non-supersymmetric theories.  One is for obtaining
some {\it qualitative} insights into strong coupling phenomena.
Another is as a testing grounds for general conjectures, e.g. the
$c$-theorem, and non-perturbative techniques.  For example instanton
technology has been checked and extended by comparing with exact
results obtained via supersymmetry~\cite{Mattis}. (Also subtleties
concerning certain exact results, as well as new exact results, were
obtained using the instanton technology discussed in~\cite{Mattis};
see references cited therein.)

One might wonder about obtaining more direct, quantitative,
information for non-supersymmetric theories by starting with a
supersymmetric theory, for which exact results can be obtained, and
perturbing by adding supersymmetry breaking terms.  This works for
small, soft, supersymmetry breaking terms $m_s$, but the $m_s/|\Lambda
|$ corrections are {\it not} under control.  There can be phase
transitions in $m_s/|\Lambda |$, which can not be avoided, i.e. a
$m_s/\Lambda$ ``wall.'' There is, in fact, evidence for such phase
transitions: the nearly supersymmetric, small $m_s$, physics is {\it
qualitatively different} from the non-supersymmetric, large $m_s$
physics in various examples~\cite{ChenShad,ArkRat}.

\section{A quick tour of the 4d, ${\cal N}=1$ susy gauge theory
landscape}

Returning to our analogy between the renormalization group and streams
of water flowing over mountains and valleys, into lakes, we start our
tour of the landscape at the bottom of a vast mountain range, with
pure ${\cal N}=1$ supersymmetric glue.  This is at the bottom of the
range because other theories, with vector-like matter, flow down to
pure glue in the infrared upon adding masses for the matter fields.

Pure ${\cal N}=1$ supersymmetric glue, with no additional matter
fields, is the same as ${\cal N}=0$ supersymmetric Yang-Mills, with
gauge group $G$, along with massless adjoint fermion matter fields
$\lambda$, which are the gluinos.  The infrared physics of these
theories is confinement, with a mass gap, and $Z_{2h}\rightarrow Z_2$
chiral symmetry breaking.  Here $h\equiv C_2(G)$ is the quadratic
Casimir of the adjoint and $Z_{2h}$ is the anomaly-free, discrete
subgroup of the $U(1)$ global $\lambda$ fermion-number symmetry which
is unbroken by instantons, which lead to $\ev{\prod _{i=1}^h(\lambda
_\alpha \lambda ^\alpha )(x_i)}=(const.)\Lambda ^{3h}$; this is
independent of the positions $x_i$ (as guaranteed by supersymmetry),
and factorization for widely separated $x_i$ suggests $\ev{\prod
_{i=1}^h(\lambda \lambda)}\rightarrow \ev{\lambda \lambda }^h$.  The
$Z_2$ is the subgroup left unbroken by gaugino condensation:
$\ev{\lambda _\alpha \lambda ^\alpha }\sim e^{2\pi ik/h}\Lambda ^3$.
Note that gaugino condensation has the quantum numbers of a
``fractional instanton'' and thus doesn't correspond to any known,
semi-classical, field configuration.  Associated with the chiral
symmetry breaking, there are $h$ supersymmetric vacua, with mass gap,
which are related by rotating the theta angle as $\theta \rightarrow
\theta +2\pi$.

A ``proof'' of the above statements follows by adding vector-like
matter, of a type so that this new theory is easier to ``solve'' than
the original, pure-glue theory.  Starting from the solved theory with
additional matter, we give the vector-like matter masses $m$.
Symmetries of the theory with added matter guarantee that the result
of this procedure is always $h$ supersymmetric vacua.  Because there
are no phase transitions in the complex mass parameter $m$,  the
pure-glue theory obtained in the $m\rightarrow \infty$ limit, where
the added matter decouples, must also have $h$ supersymmetric vacua.
This method dates back to~\cite{ShifVain}.

An old puzzle is that Witten's original calculation~\cite{WittInd} of
the index $Tr (-1)^F$, which {\it should} be the number of
supersymmetric vacua, gave $r+1$ rather than $h$, where $r$ is the
rank of the gauge group $G$.  For $SU$ and $Sp$ groups, the two
results agree, as $r+1=h$ for these cases, but for the other groups,
$SO$, $G_2$, $F_4$, $E_{6,7,8}$, the two answers disagree, as $r+1\neq
h$.  This puzzle was recently resolved by Witten~\cite{EWVS}, who
showed that the computation of $Tr (-1)^F$ can miss contributions and
verified that $Tr (-1)^F=h$ for the $SO$ groups, as well as the $SU$
and $Sp$ groups, which work as before.  The $G_2$ case was similarly
verified~\cite{Smilga}.

An unresolved puzzle~\cite{mismatch} is the normalization of
$\ev{(\lambda \lambda )^h}$.  There are two methods to compute the
normalization.  The first is a direct instanton calculation in the
strongly coupled, pure-glue, theory.  The second is to extract it from
a different instanton calculation, in a weakly coupled theory with
additional, massive, vector-like matter.  The two methods {\it
disagree} ...

A recent claim~\cite{KovShif} is that there are additional vacua, with
unbroken chiral symmetry: $\ev{\lambda \lambda}=0$. {\it If} this is
true, it would have dramatic consequences for the entire web of
interrelated theories.  All supersymmetric gauge theory results
(SQCD~\cite{Nati}, Seiberg-Witten~\cite{swi}, etc.) would need
modification.  It seems quite difficult (probably impossible) to
consistently modify {\it everything} to allow for this possibility,
and for this reason, I personally find such a chirally symmetric
vacuum to be quite unlikely (and I also find the motivation to be not
so compelling).  In any case, the subject perhaps deserves further
investigation.

It has recently been appreciated~\cite{domain} that there are domain
walls between the various supersymmetric vacua with different
$\ev{\lambda \lambda}$.  For example, for $x_3\rightarrow +\infty$,
the vacuum can be in the $\ev{\lambda \lambda}=\Lambda ^3$ vacuum
while, and for $x _3\rightarrow -\infty$ it could be in the
$\ev{\lambda \lambda}=e^{2\pi i/h}\Lambda ^3$ vacuum, with the two
phases separated by a stable domain wall.  These domain walls can
saturate a BPS bound.  Using a connection with string theory, it was
argued~\cite{EWdom} that a flux-tube string, which usually connects a
quark charge $q$ to an anti-quark charge $\overline q$, can end on the
domain walls.

We now consider some general aspects of ${\cal N}=1$ supersymmetric
theories with matter.  The landscape depends on $h=C_2(G)$ versus
$\mu =\sum _fC_2(R_f)$, where the sum is over all matter fields $f$
and $R_f$ is the representation of the gauge group which that matter
field is in.  Taking into account the one-loop beta function, the
theories are asymptotically free for $\mu <3h$.

For $\mu\geq h$, there is an {\it exactly} degenerate ``moduli space''
of physically inequivalent vacua.  Although this degeneracy is not
protected by any standard symmetry, it is ensured by holomorphy
constraint coming from supersymmetry, along with the boundary
condition that the theory behave properly at weak coupling.
Associated with the continuously degenerate vacua, there are exactly
massless moduli fields, which generally are not Goldstone bosons
(though some could be).  

For $\mu <h$, i.e. less matter than in the situation described above,
there is a classical vacuum degeneracy, similar to that described
above.  But for $\mu <h$, at the quantum level, this degeneracy is
generically lifted by non-perturbative effects, which dynamically
generate a superpotential $W_{dyn}\neq 0$.  $W_{dyn}$ can be exactly
calculated using the holomorphy constraints, along with symmetries,
connecting to known limits, and a universal, weakly coupled, $SU(2)$
instanton calculation~\cite{ADS,holom}.  Asymptotic freedom implies
that the potential associated with $W_{dyn}$ is large at small field
expectation values and slopes to zero at large expectation values.
Thus the theory with $W_{dyn}\neq 0$ actually has no stable vacuum for
$W_{tree}=0$.  By adding a $W_{tree}\neq 0$, it is possible to obtain
a stable vacuum.  In some models, this stable vacuum dynamically
breaks supersymmetry; see~\cite{PopTriv} for a recent review.

Interestingly, some special models with $\mu <h$ have inequivalent
``branches'' i.e. phases of the theory.  The branches are labeled by
a discrete quantum parameter.  This only occurs where there are {\it
no} matter fields in faithful representations of the center of the
gauge group $G$.  This is the same condition as for having {\it
distinct} Higgs, confining, or oblique confining phases: there must be
some external test charges, charged under the gauge group $G$, which
can not be screened by the dynamical matter fields.  Wilson or 't
Hooft loops involving these test charges can then have either area or
perimeter law dependence, serving as order parameters for inequivalent
phases.  The different branches correspond to the different phases.

The landscape for {\it all} theories with simple gauge group $G$ and
matter $\mu \leq h$ has now been completely charted out.  In several
recent works~\cite{GrinNolte,DottManS,CsaSki},
these theories have been comprehensively discussed, with all remaining,
previously unsolved, cases analyzed.

Theories with $\mu =h+1$ always have a quantum moduli space of vacua,
which coincides with the classical moduli space of vacua.  Often, as
in the classic~\cite{Nati} case of $SU(N_c)$ gauge theory with
$N_f=N_c+1$, the low energy theory is {\it free field theory}, with no
gauge fields, everywhere on the moduli space.  The correct variables
for the free fields are the confined meson or baryon moduli.  At the
origin, there are some additional massless confined meson or baryon
moduli fields and the sigma model metric is flat in terms of these
confined moduli.  Away from the origin, the additional massless fields
get a mass via a superpotential, which is ``dangerously irrelevant''
at the origin.  The landscape of such ``s-confining'' models based on
simple gauge groups has been systematically surveyed~\cite{CSSconf},
with previously unsolved models analyzed.

Not all models with $\mu =h+1$ are so simple.  Generally, models with
no matter fields in a faithful representation of the gauge group can
exhibit other, more interesting, types of phenomena.  For example,
$SO(N_c)$ with $N_f=N_c-1$ matter fields in the ${\bf N_c}$
representation of $SO(N_c)$ has $\mu =h+1$, but the matter is not in a
faithful representation of the gauge group, as test charges in the
spinor rep of $SO(N_c)$ test charges can not be screened by the vector
rep matter.  Rather than a theory of free chiral superfields, the
theory at the origin has a non-trivial RG fixed point for $N_c=3$;
this is an interesting theory with electric-magnetic-dyonic
triality~\cite{ISson}.  For $N_c\geq 4$, the theory at the origin has
free-magnetic, composite gauge invariance, with gauge group $SO(3)$
and $N_f=N_c-3$ flavors~\cite{ISson}.  The phases of these theories
were recently further analyzed in~\cite{Strass} by starting with a
theory with additional, vector-like, matter in a faithful
representation of the center of the gauge group, whose addition means
that distinct phases do not occur, and then decoupling this field by
giving it a large mass.

For $h+1<\mu<3h$ and no tree-level superpotential, $W_{tree}=0$, {\it
all theories} have RG fixed points or possibly free-magnetic phases
(reviewed below) at the origin of their moduli space.  The dynamics of
the theory at the origin is generally poorly understood, and the
landscape remains largely uncharted.

Some interesting phenomena have been observed in a hodge-podge of
examples, with no general understanding.  One is that two different
looking theories, with different gauge groups and matter fields, can
flow to the same renormalization group fixed point.  At the fixed
point, both give {\it exactly} the same physics.  The original example
of this for 4d asymptotically free (as opposed to finite) theories is
Seiberg's SQCD duality~\cite{Nati}, between $SU(N_c)$ with $N_f$
fundamental flavors and $SU(N_f-N_c)$ with $N_f$ flavors, some singlet
fields to be identified as the mesons of the original theory, and a
superpotential.  For $3N_c>N_f>\f{3}{2}N_c$, both theories flow to the
same, interacting, renormalization group fixed point.  The $SU(N_c)$
theory is scale invariant at some coupling constant $g^*_{N_c,N_f}$
and the dual $SU(N_f-N_c)$ is the same scale invariant theory at some
coupling $\widetilde g^*_{N_f-N_c,N_f}$.  Which description of the
fixed point is more useful depends on which $g^*$ is small.
Generally, smaller $g^*$ must correspond to larger $\widetilde g^*$,
though the precise map between the two is not known.  It is worth
emphasizing that duality is inherently quantum mechanical: the two
dual theories are completely inequivalent at the classical level.

An entirely new phenomenon, which was also discovered in Seiberg's
seminal paper on duality~\cite{Nati}, is the existence of a
non-Abelian, free magnetic phase in four dimensional ${\cal N}=1$
supersymmetric theories.  There are low-energy fields, which are
essentially solitons of the UV theory, which behave as quarks and
gluons of a non-Abelian gauge theory which is IR free.  The magnetic
quarks and gluons are the solution for the low energy spectrum.  The
composite gluons show that gauge invariance does not have to be
fundamental -- there can be composite gauge invariance.

There are many examples of duality and free-magnetic phases, all found
via some guess-work and many non-trivial cross-checks, fitting into a
growing web of interrelated examples.  There is no known general
criteria which can be generally applied to determine whether a given
theory has an interacting RG fixed point, a free-magnetic phase, or
something else.  At present, one has to work on a case-by-case,
basis.  It is also not generally known, if a theory does have a RG
fixed point, whether it should have a dual description, and what that
dual description should be.  There are still many confusing examples
which remain unsolved, and all examples should be understood at a
deeper level.  

For $SU$, $SO$, and $Sp$ groups $G$ with (only) fundamental matter,
the duality has been ``derived'' by renormalization group flows from
their ${\cal N}=2$ supersymmetric analogs~\cite{APS}.  Using exact
results in the ${\cal N}=2$ supersymmetric theories, the dual gauge
group $\widetilde G$ is infrared free, and its gauge group and matter
content is directly seen.  Breaking to ${\cal N}=1$ supersymmetry at
scale $m_s$, we have two different flows.  The first has $m_s\gg
\Lambda$ and flows first very close to the ${\cal N}=1$ theory with
gauge group $G$, and eventually to the fixed point of that theory.  The
second type of flow occurs for $m_s\ll \Lambda$.  Then the dynamics
starts off controlled by the ${\cal N}=2$ theory, and the theory first
flows very close to the ${\cal N}=2$ theory with dual gauge group
$\widetilde G$ and matter content.  Eventually, the ${\cal
N}=2\rightarrow {\cal N}=1$ breaking due to $m_s$ kicks in, and the
theory flows close to the ${\cal N}=1$ supersymmetric theory with dual
gauge group $\widetilde G$ and matter content.  Eventually, that
theory flows to some fixed point.  Now, assuming that the two
different flows really are close to the two dual ${\cal N}=1$
theories, and since ${\cal N}=1$ supersymmetry should prohibit phase
transitions in $m_s/\Lambda$, the two dual theories must, in fact,
flow to the same fixed point.

This ``proof'' has a direct analog in the recent brane constructions
of $4d$ gauge theories~\cite{EGK}.  However, this requires several
more assumptions about the dynamics of branes and string theory, so
perhaps this is better referred to as a ``relation'' than a ``proof.''

\section{Results in other dimensions and connections with string theory}

There have also been a variety of results for a variety of gauge
theories in other dimensions.  Note that for $d\neq 4$ the gauge
coupling is dimensionful.  The effective, dimensionless, gauge
coupling at an energy scale $E$ is $g_{eff}=gE^{(d-4)/2}$.  Because of
this classical scale dependence, all gauge theories are asymptotically
free for $d<4$ and infra-red free (i.e. ``nonrenormalizable'') for
$d>4$.  

For $d=3$, there are many interesting RG fixed points with various
supersymmetries and dualities.  As an example with ${\cal N}=2$
supersymmetry in 3d (this has the same number of supercharges as
${\cal N}=1$ in 4d), a Wess-Zumino theory with a single chiral
superfield $X$ and superpotential $W=X^3$ flows to an interacting RG
fixed point.  As another set of examples, SQED, with $U(1)$ gauge
group and $N_f>0$ flavor of fields with charges $\pm 1$ flow to RG
fixed points.  For the case $N_f=1$, this SQED fixed point has a dual
description in terms of a Wess-Zumino theory with chiral superfields
$X$, $Y$, and $Z$, with superpotential $W=XYZ$~\cite{AIS}.  There are
also ``mirror symmetry'' dual descriptions of RG fixed points with
${\cal N}=4$ supersymmetry in 3d~\cite{ISmir}; this duality exchanges:
Higgs $\leftrightarrow$ Coulomb branches, classical $\leftrightarrow$
quantum, masses $\leftrightarrow$ Fayet-Iliopoulos terms, and manifest
global symmetries $\leftrightarrow$ hidden quantum symmetries.

More surprisingly, it has been found (via string theory) that
non-trivial RG fixed points exist for various (supersymmetric) field
theories in $d=5$ and $d=6$.  For $d=5$ there is a complete
classification of all supersymmetric gauge theories which {\it exist}
via flows from 5d RG fixed points~\cite{5dref}.  The fixed points
themselves are not well understood.  Perturbing them by a relevant
operator, which corresponds to the 5d gauge kinetic term, they flow to
particular, IR free, 5d gauge theories which, in this sense,
``exist.'' For example, for $SU(N_c)$ gauge group, the theory exists
for $N_f\leq 2N_c$ flavors; this is the analog of the constraint for
asymptotic freedom in 4d.  In 5d, RG fixed points only exist with
minimal ${\cal N}=1$ supersymmetry~\cite{Nahm,Minwalla} (which has the
same number of supercharges as ${\cal N}=2$ in 4d).

In $d=6$ there can be chiral supercharges and non-trivial RG fixed
points exist for the minimal ${\cal N}=(1,0)$
supersymmetry~\cite{SW6d,NS6d,KI6d} (which has the same number of
supersymmetries as ${\cal N}=2$ in 4d) and for ${\cal N}=(2,0)$
supersymmetry.  The ${\cal N}=(1,1)$ theories are necessarily
free~\cite{Nahm,Minwalla} and theories with higher ${\cal N}$ have
fields with spins up to two, i.e. necessarily include gravity.  The
${\cal N}=(2,0)$ theories do not include standard gauge fields, which
are not allowed by the $(2,0)$ supersymmetry, but rather chiral 2-form
gauge fields, with self-dual field strengths: $A_{\mu \nu}$, with
$dA=*dA$.  The ${\cal N}=(1,0)$ theories can include gauge fields, but
are either free or anomalous unless self-dual, two-form gauge fields
$A_{\mu \nu}$ are also present.

In all cases with 6d non-trivial fixed points, there are BPS strings,
which appear to become tensionless at the origin of the ``Coulomb
branch,'' where scalar moduli partners of $A_{\mu \nu}$ vanish.
Nevertheless, it seems possible to interpret the theory at the origin
as in interacting, 6d RG fixed point {\it field
theory}~\cite{LSunp,SW6d} rather than requiring some new and unknown
kind of ``tensionless string theory.''

The ${\cal N}=(2,0)$ theory can be constructed from type IIB string
theory compactified down to 6d on a space which is allowed to become
singular~\cite{Wcomments} or in IIA string theory or M-theory via
branes~\cite{Stropen}.  The world-volume of $N_c$ type IIA or $M$
theory 5-branes has a 2-form version of $SU(N_c)$, with $N_c$ abelian
2-forms $A_{\mu \nu}^a$, with $dA=*dA$, corresponding to the Cartan of
the $SU(N_c)$ and 6d strings corresponding to the
$W$-bosons~\cite{Wcomments}.  At the origin of the moduli space,
where the strings appear to become tensionless, there is the
interacting ${\cal N}=(2,0)$ RG fixed point conformal field theory.

The interacting 6d ${\cal N}=(2,0)$ theory gives an ordinary gauge
theory when reduced to $d<6$ on a circle.  Going to 4d by making two
directions circles of radii $R_1$ and $R_2$ yields a 4d, ${\cal N}=4$
$SU(N_c)$ gauge theory with $g_{YM}^2=R_1/R_2$.  Because we can
obviously exchange the names of the two circles, this construction
makes the $g_{YM}\leftrightarrow 1/g_{YM}$ Montonen-Olive duality of
this theory manifest.  It is also possible to obtain a theory similar
to 4d, non-supersymmetric, QCD from the 6d theory by changing the
boundary conditions on the circles.  The 4d, non-supersymmetric theory
thus obtained is referred to as ``MQCD'' and it is hoped to be in the
same universality class as real-world QCD, without phase
transitions~\cite{EWdom}.

There has been a fruitful interplay between gauge theories and string
theories over the past few years.  It is possible to get composite
gauge invariance in string theory in a variety of (related) ways:
singular compactification geometry, zero size instantons, and branes.
There is a correspondence between results in the field theory thus
obtained and results concerning string theory.  Generally the
correspondence between field theory and string theory results involve
opposite limits: where one side is known well (perhaps weakly
coupled), the other side is often poorly understood.  An example of a
nice interplay between field theory and string theory
is~\cite{SenBDS}. In this way, known results on one side translate
into new results on the other side or, in the case where both sides
are understood, cross-checks of the correspondence are obtained.

An example by which composite gauge invariance is obtained in string
theory is via $D$-branes, which can be thought of as being similar to
solitons of string theory.  These theories have supersymmetric gauge
theories living in their world-volume; see~\cite{JPTasi} for an
extensive review with references.  A simple example is $N_c$ parallel
D3 branes of type $IIB$ string theory, which has 4d, ${\cal N}=4$,
$SU(N_c)$ gauge theory living in its world-volume.  (I ignore the
subtle issue about if this theory is $U(N_c)$ or $SU(N_c)$.)  Various
other 4d (and other $d$) theories, with fewer supersymmetries, can be
obtained via more complicated arrangements of branes.  It is possible
to use known field theory results to obtain the rules governing
$D$-branes.  Knowing these rules, along with various string dualities,
$D$-branes prove to be a powerful tool to generalize to new examples
and obtain new information about field theory.  See~\cite{GKrev} for
an extensive review, with references.

\section{Renormalization group fixed points and $AdS$.}

The Maldacena conjecture~\cite{malda,GKP,EWh}, in the generalized
sense of~\cite{EWh}, is that a gravity theory in $d+1$ dimensional
anti de Sitter space, $AdS_{d+1}$, is {\it dual} to a $d$ dimensional
conformal field theory on the boundary of $AdS_{d+1}$.  $d+1$
dimensional anti de Sitter space is a solution of Einstein's equations
with negative cosmological constant, $\Lambda <0$, and has a time-like
boundary, which is ordinary $d$-dimensional Minkowski spacetime.  The
extra space dimension associated with the bulk of the $d+1$
dimensional $AdS_{d+1}$ is to be thought of as roughly the
renormalization group parameter of the boundary field theory.  

A motivation for such a duality is that the $d$ dimensional conformal
group is the same as the symmetry group of $AdS_{d+1}$.  This symmetry
group is inherited by the conformal field theory on the boundary of
$AdS_{d+1}$.  A concrete relation between the two dual theories
is~\cite{EWh}
$${\cal Z}_{gravity}\left[\Phi _i|_{\partial (AdS)}=J_i(x)\right] =
\ev{e^{\sum _i \int d^dx J_i(x){\cal O} _i(x)}}_{CFT},$$ where $\Phi
_i$ are $AdS_{d+1}$ gravity fields, ${\cal O}_i$ are associated CFT
operators, and $J_i(x)$ are arbitrary source functions.  There is, in
this way, a map between all operators ${\cal O}_i(x)$ of the $d$
dimensional boundary conformal field theory and the fields $\Phi _i$
of the gravity theory.

This duality realizes~\cite{EWh} the ``holography'' of gravity, due to
't Hooft, Thorn, and Susskind~\cite{thoofth,thorn,Suss}.  The
physics of the $d+1$ dimensional, bulk, gravity theory is encoded in
that of a $d$ dimensional boundary field theory.  In a theory of
gravity, one dimension can be regarded as a holographic illusion.  

The original example of this duality was obtained~\cite{malda} 
by considering $D$-branes vs. the throat geometry of the associated
black holes which carry the same quantum numbers.  In this way, it was
argued that 4d ${\cal N}=4$,  $SU(N_c)$ super-Yang-Mills theory is dual
to type IIB string theory on $AdS_5\times S^5$, with $N_c$ units of
$F_5$ flux on the $S^5$.  The string description is weakly coupled for
the limit of small $g_{YM}$, with $\lambda =g_{YM}^2N_c$ large.  The
Yang-Mills theory, on the other hand, is weakly coupled in the limit
of $\lambda$ small.  Thus, as is always the case, where one
description of the physics is weakly coupled, the dual description is
strongly coupled.  Here the gravity or string theory can be regarded
as the large $N_c$ master-field for $\lambda \gg 1$.  

The duality has been generalized to ${\cal N}=4$ theories with $SO(N)$
and $Sp(N)$ gauge groups~\cite{EWbar} and to theories with ${\cal
N}=2,1,0$ supersymmetry~\cite{SKESLNV}.  Because the $AdS$ space
remains untouched in these constructions, the resulting theories with
${\cal N}=2,1,0$ susy also have a line of RG fixed points, i.e. are
finite theories with $\beta (g)\equiv 0$, at least in the limit of
large $N_c$.  It is also possible to break supersymmetry via finite
temperature $T$, going from the 4d supersymmetric theory to a 3d
non-supersymmetric theory by putting the theory on a Euclidean circle.
It is also possible to obtain in this way the 4d non-supersymmetric
MQCD via the 6d theory with ${\cal N}=(2,0)$
supersymmetry~\cite{EWii}, which is dual~\cite{malda} to $M$ theory on
$AdS_7\times S^4$.

Wilson loop correlation functions are computed in the $AdS$ dual
via~\cite{Mhooft}
$$\ev{\prod _i W(C_i)}\sim e^{-Area[S(C_i)]},$$ where $S(C_i)$ is the
minimal area, 2d world-sheet living in the $d+1$ dimensional bulk
whose boundary is $\oplus C_i$, the Wilson loops $C_i$ living in the
$d$ dimensional boundary.  In the 4d ${\cal N}=4$ theory conformal
phase, in the limit $g_{YM}\rightarrow 0$ with $g_{YM}^2N$ large,
where the gravity dual is weakly coupled, this gives~\cite{Mhooft} for
the potential between two test charge sources separated by distance
$R$:
$$V(R)=-{4\pi ^2\over \Gamma (\f{1}{4})^4}{\sqrt {g_{YM}^2N}\over
R}.$$ Note that the $g_{YM}$ dependence differs from the $V(R)\sim
g^2/R$ expected for small $g_{YM}^2N$; the above result is thus
interpreted as a non-trivial, new prediction for the theory in the
limit of large $g_{YM}^2N$.  Presumably, there is some function of
$g_{YM}$ and $N$ which interpolates between the two results.  

Applying to 4d non-susy MQCD, obtained from higher dimensions via
finite $T$, the potential is found to be~\cite{sigr}
$$V(R)=\sigma R +O(e^{-RT}), \qquad \sigma \sim (g_{YM}^2N)T^2.$$ Also
glueball masses have been analyzed in this limit~\cite{glueball}.  A
pecular result, as emphasized by~\cite{GrenOle}, is that
$M_{glueball}$ is $g_{YM}$ independent, so the above expression for
the string tension implies that $M_{glueball}/\sqrt{\sigma}
\rightarrow 0$ in the large $g_{YM}^2N$ limit.         

\section{Summary}

To summarize, there are lots of predictions for strong coupling.  It
would be nice to have a deeper understanding, and also non-perturbative
methods to directly check them!  This is a challenge for the future.

\section*{Acknowledgments}
I would like to thank the organizers for inviting me to speak at this
interesting conference.  I am especially grateful for their extensive
assistance and patience, and for the many friendly reminders to
complete this contribution.  I am supported by UCSD DOE grant
DOE-FG03-97ER40546 and an Alfred Sloan Foundation Fellowship.  
This contribution was written while I was a visitor at the IAS, and
supported in part by the W.M. Keck Foundation.

\end{document}